\documentclass[preprint,
showpacs,
amsmath,amssymb,
aps
]{revtex4-1}
\usepackage{dcolumn}
\usepackage{bm}
\usepackage{mathrsfs}
\usepackage[titletoc]{appendix}

\begin{document}

\title{New Energy-Momentum and Angular Momentum Tensors with Applications to Nucleon Structure}
\author{Zhen-Lai Wang}
\author{De-Tian Yang}
\author{Zi-Wei Chen}
\author{Tao Lei}
\author{Xiang-Song Chen}%
\email{cxs@hust.edu.cn}
\affiliation{School of Physics and
MOE Key Laboratory of Fundamental Quantities Measurement,
Huazhong University of Science and Technology, Wuhan 430074, China}

\date{\today}
\begin{abstract}
We present a new type of energy-momentum tensor and angular momentum tensor. They are motivated by a special 
consideration in quantum measurement: Given a wave in mutual eigen-state of more than one physical observables, 
the corresponding physical currents should be proportional to each other.  Interestingly, this criterion denies the 
traditional canonical and symmetric expressions of energy-momentum tensor and their associated expressions of 
angular momentum tensor. The new tensors we propose can be derived as N\"other currents from a Lagrangian with 
second derivative, and shed new light on the study of nucleon structures.   
\end{abstract}

\maketitle

\section{Three types of energy-momentum tensor}

Being the conserved currents associated with the symmetries of space-time translation and rotation, energy-
momentum and angular momentum tensors are among the most fundamental quantities in both classical and 
quantum physics, and are the basis for obtaining the structural picture of a physical system, e.g., the quark-gluon
structure of nucleon spin, which remains an open problem \cite{Lead14}. 
There are two popular expressions of energy-momentum tensor, the canonical one and the symmetric one, each has 
a corresponding expression of angular momentum tensor. In this paper, we present a new type of energy-momentum 
and angular momentum tensors, and apply them to the study of nucleon structures.  To show the difference, let us first 
give the explicit expression of our new energy-momentum tensor. For a free field $\phi_a$, it is 
\begin{equation}
T_{\rm new}^{\mu\nu}=-\dfrac{\partial{\mathscr  L}_{\rm st}}{\partial(\partial_\mu\phi_a)}
\overleftrightarrow{\partial}^\nu\phi_a,
~\overleftrightarrow{\partial}^\nu=\dfrac{1}{2}(\overrightarrow{\partial}^\nu-\overleftarrow{\partial}^\nu).\label{Tnew}
\end{equation}
Here, ${\mathscr  L}_{\rm st}(\phi_a,\partial_\mu\phi_a)$ is the conventional expression of Lagrangian in terms of the 
field $\phi_a$ and its first derivative. We take the metric with signature $(-+++)$.
In comparison, the canonical expression is 
\begin{equation}
T_{\rm cano}^{\mu\nu}=-\dfrac{\partial{\mathscr  L}_{\rm st}}{\partial(\partial_\mu\phi_a)}{\partial}^\nu\phi_a 
+g^{\mu\nu}{\mathscr  L}_{\rm st}.\label{Tcano}
\end{equation}
For example, for the most familiar electromagnetic field with Lagrangian ${\mathscr L}_{\rm st}^{A}=
-\dfrac{1}{4}F_{\mu\nu} F^{\mu\nu}$, the three types of energy-momentum tensor read:
\begin{subequations}
\begin{align}
^A T^{\mu\nu}_{\rm new}&=F^{\mu\rho}\overleftrightarrow{\partial}^{\nu}A_\rho
=\frac{1}{2}(F^{\mu\rho}\partial^{\nu}A_\rho-A_\rho \partial^{\nu}F^{\mu\rho}) ,\label{Anew}\\
^A T^{\mu\nu}_{\rm cano}&=F^{\mu\rho} \partial ^{\nu}A_\rho +g^{\mu\nu}{\mathscr  L}_{\rm st}, \label{Acano}\\
^A T^{\mu\nu}_{\rm symm}&=F^{\mu\rho} F ^\nu_{~\rho} +g^{\mu\nu}{\mathscr  L}_{\rm st} \label{Asymm}.
\end{align}
\end{subequations}

\section{Justification of the new expression}
Our motivation to re-examine the expressions of energy-momentum and angular momentum tensors is that 
the conserved currents are not uniquely determined by the conservation laws, which can only prescribe total 
conserved charges; and much debate arose, especially in the attempt to decomposing the nucleon spin 
into the spin and orbital contributions of  quarks and gluons 
(see Refs. \cite{Lead14, Lead16, Waka16} for most recent discussions).
On the other hand, the conserved current
{\em densities} do have independent physical meanings. The most familiar example is that the energy-momentum 
tensor acts as the source of gravitational field in Einstein's gravitational theory. The less known spin current, in 
Einstein-Cartan theory \cite{Hehl76}, is taken as the source of space-time
torsion.  In either the Einstein theory or the Einstein-Cartan theory, however, the energy-momentum tensor and the 
spin tensor are derived {\it post  priori} from the action constructed with minimal coupling. In this paper, we seek to set  
{\em a priori}  constraint on the energy-momentum and angular momentum tensors.  

Our consideration is that \textit{if a quantum wave is in mutual eigen-state of more than one physical observables, 
and a simultaneous measurement of these observables can be performed, then the currents associated with these 
observables must be proportional to each other.}

The hint on such correlation of currents comes from classical particles: When one catches a classical particle, one 
catches all its physical quantities: charge, energy, momentum, etc. Thus, for a beam of identical particles with the 
same energy $\varepsilon$ and momentum $p_j$ for each particle, the  energy flux density $\vec {\cal K}_0$ must be 
proportional to the momentum flux density $\vec {\cal K}_j$:
\begin{equation}\label{classical}
\dfrac {\vec {\cal K}_0}{\varepsilon}= \dfrac {\vec {\cal K}_j}{p_j} =\vec {\cal K}_n,
\end{equation}
with $\vec {\cal K}_n$ the flux density of particle number.
[The case will be trivial if all components of $p_j$ are identical for the particles, but non-trivial cases 
can be designed if just one or two components of $p_j$ are set identical. The same remark applies to the discussion 
of quantum wave below.]

By the assumption of quantum measurement, when a quantum wave collapses to a local spot, all its physical 
quantities will localize simultaneously to that same spot. In this way, a quantum wave should exhibit similar correlation 
of currents as for classical particles: If the wave is in mutual eigen-state of energy $\varepsilon$ and momentum 
$p_j$, then the density of energy flow $T^{i0}$ and the density of momentum flow $ T^{ij}$ must satisfy 
a constraint similar to Eq. (\ref{classical}):
\begin{equation}\label{quantum}
\dfrac {T^{i0}}{\varepsilon}= \dfrac{T^{ij}}{p_j},
\end{equation}
so that one can have 
\begin{equation}\label{measure}
\dfrac {T^{i0}\cdot dS_i}{\varepsilon}= \dfrac{T^{ij}\cdot dS_i}{p_j}=\dfrac{dN}{dt},
\end{equation}
where $N$ is the number of particles received at the surface element $d\vec S$.

It is interesting and surprising that the conventional expressions of energy-momentum tensor do not show the 
above correlation. 
For example, taking the canonical expression in Eq. (\ref{Tcano}), and making use of the eigen-state assumption 
$\partial^0\phi_a=i \varepsilon \phi_a$,  
$\partial^j\phi_a=i p_j \phi_a$, we have:  
\begin{subequations}
\begin{align}
T_{\rm cano}^{i0}&\rightarrow- i \varepsilon\dfrac{\partial{\mathscr  L}_{\rm st}}{\partial(\partial_i\phi_a)}\phi_a, 
\label{i0} \\
T_{\rm cano}^{ij}&\rightarrow- i p_j\dfrac{\partial{\mathscr  L}_{\rm st}}{\partial(\partial_i\phi_a)}\phi_a+
\delta_{ij}{\mathscr  L}_{\rm st}.\label{ij}
\end{align}
\end{subequations}
This satisfies the constraint in Eq. (\ref{quantum}) for the transverse momentum flow, namely $T^{ij}$ with $i\neq j$. 
But for the longitudinal momentum flow $T^{jj}$,  the Lagrangian term in Eq. (\ref{ij}) makes a trouble for Eq. 
(\ref{quantum}), except for the Dirac field $\psi$ which has 
$ {\mathscr  L}^\psi_{\rm st}= 0$ when applying the equation of motion. 

Such a Lagrangian term also exists in the symmetric expression of energy-momentum tensor, which therefore does 
not fulfill the constraint in Eq. (\ref{quantum}), either. In fact, the symmetric energy-momentum tensor stands an even 
worse situation with respect to such a constraint: 
One can check with the familiar electromagnetic field that $^AT^{\mu\nu}_{\rm symm}$ in Eq. (\ref{Asymm}) does not 
guarantee Eq. (\ref{quantum}) even for $i\neq j$.

Our new expression of energy-momentum tensor in Eq. (\ref{Tnew}) does not contain the Lagrangian term. 
In the next section, we derive an equivalent and more illuminating expression than Eq. (\ref{Tnew})
to display that such current-correlation property
can be safely guaranteed for a quantum wave in mutual eigen-state of energy $\varepsilon$ and momentum $p_j$.

\section{Derivation as N\"other current}

The  conventional canonical energy-momentum tensor is derived as a N\"other current with the conventional 
Lagrangian ${\mathscr  L}_{\rm st}(\phi_a,\partial_\mu\phi_a)$. From our above discussion, we see that it 
almost satisfies the constraint in Eq. (\ref{quantum}), except for the 
Lagrangian term which does not in general vanish. Since the Lagrangian of a field can be modified by a surface term 
without changing the equation
of motion, this gives us a hint that if we can find a general expression of Lagrangian which always vanishes after 
applying the equation of motion, 
then the derived N\"other current will automatically satisfy the constraint in Eq. (\ref{quantum}). In what follows, we 
show that it is indeed so. We will first concentrate on the free-field case which is already non-trivial, and discuss the  
interacting case in section \ref{nucleon}, in connection with the quark-gluon structure of nucleon.

The conventional standard Lagrangians of free scalar, Dirac, and vector fields take the following forms, respectively:
 \begin{subequations}\label{Lst}
\begin{align}
{\mathscr L}_{\rm st}^{\phi}&=-\dfrac{1}{2}\partial^\mu\phi\partial_\mu\phi-\dfrac{1}{2}m^2\phi^2 ,\\
{\mathscr L}_{\rm st}^{\psi}&=\dfrac{1}{2}\overline{\psi}(i\gamma^{\mu}\partial_{\mu}-m)\psi+h.c ,\\
{\mathscr L}_{\rm st}^{A}&=-\dfrac{1}{4}F_{\mu\nu} F^{\mu\nu}.
\end{align}
\end{subequations}
By noticing that a free-field Lagrangian $ {\mathscr  L}_{\rm st}(\phi_a,\partial_\mu\phi_a)$ is necessarily quadratic in 
the field variable and its derivative, it can be put in a unified form:
 \begin{equation}\label{Lquad}
 {\mathscr  L}_{\rm st}(\phi_a,\partial_\mu\phi_a)=\dfrac{1}{2}\Big[\phi_a\dfrac{\partial{\mathscr  L}_{\rm st}}
{\partial\phi_a}
+(\partial_\mu\phi_a)\dfrac{\partial{\mathscr  L}_{\rm st}}{\partial(\partial_\mu\phi_a)}\Big].
 \end{equation}
By adding a proper surface term, we obtain the desired new expression of Lagrangian 
${\mathscr  L}_{\rm new}(\phi_a,\partial_\mu\phi_a,\partial_\mu\partial_\nu\phi_a)$:
\begin{subequations}
\begin{align}
{\mathscr  L}_{\rm new}&={\mathscr L}_{\rm st}-\dfrac{1}{2}\partial_\mu\big[\phi_a\dfrac{\partial{\mathscr  L}_{\rm st}}
{\partial(\partial_\mu\phi_a)}\big] \label{surface}\\
&=\dfrac{1}{2}\phi_a\Big[\dfrac{\partial{\mathscr L}_{\rm st}}{\partial\phi_a}
-\partial_\mu\dfrac{\partial{\mathscr L}_{\rm st}}{\partial(\partial_\mu\phi_a)}\Big],\label{Lnew}
\end{align}
\end{subequations}
which clearly vanishes by the Euler-Lagrange equation of motion. 

The explicit forms of our new Lagrangian for the scalar, Dirac, and vector fields are:
 \begin{subequations}
\begin{align}
\mathscr{L}_{\rm new}^{\phi}&=\frac{1}{2}\phi(\partial_\mu\partial^\mu-m^2)\phi ,\\
\mathscr{L}_{\rm new}^{\psi}&=\frac{1}{2}\overline{\psi}(i\gamma^\mu\partial_\mu-m)\psi+h.c., \\
\mathscr{L}_{\rm new}^{A}&=\frac{1}{2}A_\nu \partial_\mu F^{\mu\nu}.
\end{align}
\end{subequations}
For the Dirac field, the ``new'' Lagrangian actually equals the traditional expression, which is already zero 
by the equation of motion.  

Notice that the new Lagrangian ${\mathscr  L}_{\rm new}$ contains a second derivative,
thus the derivation of N\"other current is a little bit (but not much) more involved \cite{Lomp14}. The result is: 
 \begin{equation}\label{TnewH}
T^{\mu\nu}_{\rm new}=-i\big[\frac{\partial{\mathscr  L}_{\rm new}}{\partial(\partial_{\mu}\phi_{a})}+\frac{\partial{\mathscr  
L}_{\rm new}}{\partial(\partial_{\mu}\partial_{\sigma}\phi_{a})}\partial_{\sigma}-\partial_{\sigma}\frac{\partial{\mathscr  
L}_{\rm new}}{\partial(\partial_{\sigma}\partial_{\mu}\phi_{a})}\big]{\bm P}^ {\nu} \phi_{a},
\end{equation}
where $\bm{P}^{\nu}=-i\partial^{\nu}$ is the quantum-mechanical four-momentum operator. 
We call this a ``hyper-canonical'' form, as it is a single expression with the single operator inserted for the 
desired observable. Such a hyper-canonical form clearly guarantees the current-correlation property as we 
elaborated above for a quantum wave in mutual eigen-state of two or more components of $\bm{P}^\nu$. 

By a slight algebra, Eq. (\ref{TnewH}) can be converted into the more convenient expression in Eq. (\ref{Tnew}), 
expressed with the 
conventional Lagrangian ${\mathscr L}_{\rm st}$ containing only the first derivative. 

As a cross-check, the difference between $T_{\rm new}^{\mu\nu}$ and $T_{\rm cano}^{\mu\nu}$ is a total-divergence 
term:
\begin{subequations}
\begin{align}
T_{\rm new}^{\mu\nu}&=T_{\rm cano}^{\mu\nu}+\partial_{\lambda}{\mathscr K}^{[\lambda\mu]\nu}, \\
{\mathscr K}^{[\lambda\mu]\nu}&=\dfrac{1}{2}\big(g^{\lambda\nu}\dfrac{\partial{\mathscr  L}_{\rm st}}
{\partial(\partial_{\mu}\phi_a)}-g^{\mu\nu}\dfrac{\partial{\mathscr  L}_{\rm st}}{\partial(\partial_{\lambda}\phi_a)}
\big)\phi_a.\label{K}
\end{align}
\end{subequations}
Here, ${\mathscr K}^{[\lambda\mu]\nu}$ is antisymmetric in its first two indices. As a consequence, 
$T_{\rm new}^{\mu\nu}$ satisfies the same conservation law and gives the same conserved four-momentum as 
does by $T_{\rm cano}^{\mu\nu}$:
\begin{subequations}
\begin{align}
&\partial_{\mu} T_{\rm cano}^{\mu\nu}=\partial_{\mu} T_{\rm new}^{\mu\nu}=0,\\
&P^{\nu}=\int{\mathrm{d}}^3 x\,T_{\rm cano}^{0 \nu }=\int{\mathrm{d}}^3 x\,T_{\rm new}^{0 \nu }.
\end{align}
\end{subequations}

\section{The angular momentum tensor}
Due to the vanishing of the new Lagrangian 
${\mathscr  L}_{\rm new}$ under
the equation of motion, the ``hyper-canonical'' structure in Eq. (\ref{TnewH}) actually applies to any N\"other current.  
The more non-trivial example is the new angular momentum tensor:
\begin{equation}\label{MnewH}
M^{\lambda\mu\nu}_{\rm new}=-i\big[\frac{\partial{\mathscr  L}_{\rm new}}{\partial(\partial_{\lambda}\phi_{a})}+
\frac{\partial{\mathscr  L}_{\rm new}}{\partial(\partial_{\lambda}\partial_{\sigma}\phi_{a})}\partial_{\sigma}-
\partial_{\sigma}\frac{\partial{\mathscr  L}_{\rm new}}{\partial(\partial_{\sigma}\partial_{\lambda}\phi_{a})}\big]{\bm J}^ 
{\mu\nu}_{ab} \phi_{b}.
\end{equation}
Here, ${\bm J}^ {\mu\nu}_{ab}$ is the total angular momentum operator that governs the transformation of $\phi_a$ 
under an infinitesimal Lorentz transformation $x_{\mu}\to x'_{\mu}= x_{\mu}+ \omega_{\mu\nu} x^{\nu}$: 
\begin{equation}\label{J}
\delta\phi_{a}(x)=\frac{1}{2} \omega_{\mu\nu}[(x^{\mu}\partial^{\nu}-x^{\nu}\partial^{\mu})\delta_{ab}+i\bm{S}
^{\mu\nu}_{ab}]\phi_{b}(x)
\equiv\frac{1}{2} \omega_{\mu\nu}i{\bm J}^{\mu\nu}_{ab}\phi_{b}.
\end{equation}
Namely, ${\bm J}^{\mu\nu}=(x^{\mu}{\bm P}^{\nu}-x^{\nu}{\bm P}^{\mu})+{\bm S}^{\mu\nu}={\bm L}^{\mu\nu}+{\bm S}
^{\mu\nu}$. It relates to 
the usual angular momentum operator by ${\bm J}_i=\frac 12 \epsilon_{ijk}{\bm J}^{jk}$. 
The same hyper-canonical structure in Eqs. (\ref{TnewH}) and (\ref{MnewH}) guarantees that if the wave 
$\phi_a$ is a mutual eigen-state 
of energy and angular momentum, say, ${\bm J}^{12}_{ab} \phi_b=j_3 \phi_a$, then the fluxes of energy and angular 
momentum will be correlated:
\begin{equation}\label{TM}
\dfrac {T^{i0}}{\varepsilon}= \dfrac{M^{i12}}{j_3}.
\end{equation}

In comparison, the canonical angular momentum tensor, 
\begin{equation}
M^{\lambda\mu\nu}_{\rm cano}=x^{\mu}T^{\lambda\nu}_{\rm cano}-x^{\nu} T^{\lambda\mu}_{\rm cano}-i\dfrac{\partial 
{\mathscr L}_{\rm st}}{\partial (\partial_{\lambda} \phi_a)}{\bm S}^{\mu\nu}_{ab}\phi_b, \label{Mcano}
\end{equation}
cannot be put in a form in which the angular momentum operator ${\bm J}^ {\mu\nu}_{ab}$ appears as a whole, and 
thus does not satisfy 
the quantum constraint as we put above. [Specifically, $M^{\lambda\mu\nu}_{\rm cano}$ contradicts
Eq. (\ref{TM}) for the transverse angular momentum flux with $i=1$ or $2$; only the longitudinal flux $M^{312}_{\rm 
cano}$ satisfies Eq. (\ref{TM}).]
It is easy to check that the angular momentum tensor associated with the symmetric energy-momentum tensor,
\begin{equation}
M^{\lambda\mu\nu}_{\rm symm}=x^{\mu}T^{\lambda\nu}_{\rm symm}-x^{\nu} T^{\lambda\mu}_{\rm symm}, 
\label{Msymm}
\end{equation}
stands a even worse situation with respect to Eq. (\ref{TM}). 

To make a more detailed comparison with $M^{\lambda\mu\nu}_{\rm cano}$, we convert $M^{\lambda\mu\nu}_{\rm 
new}$
in Eq. (\ref{MnewH}) into a more conventional form:
\begin{equation}\label{Mnew}
M^{\lambda\mu\nu}_{\rm new}=x^{\mu}T^{\lambda\nu}_{\rm new}-x^{\nu}T^{\lambda\mu}_{\rm new}-i \dfrac{\partial 
{\mathscr L}_{\rm st}}{\partial (\partial_{\lambda} \phi_a)}{\bm S}^{\mu\nu}_{ab}\phi_b
+\frac{1}{2}\big[g^{\lambda\mu}\dfrac{\partial {\mathscr  L}_{\rm st}}{\partial (\partial_{\nu} \phi_a)}-g^{\lambda\nu}
\dfrac{\partial {\mathscr L}_{\rm st}}{\partial (\partial_{\mu} \phi_a)}\big]\phi_a. 
 \end{equation}
By this form, it can be checked that the difference between $M_{\rm new}^{ \lambda \mu \nu}$ and $M_{\rm cano}
^{ \lambda \mu \nu}$ is again a total-divergence term:
 \begin{equation}
M_{\rm new}^{\lambda\mu\nu}=M_{\rm cano}^{\lambda\mu\nu}+\partial_{\rho}\big(x^{\mu}{\mathscr K}^{ [\rho\lambda ] 
\nu}-x^{\nu}{\mathscr K}^{ [\rho\lambda ] \mu}\big),
\end{equation}
with the same $ {\mathscr K}^{ [\rho\lambda ] \nu}$ as in Eq. (\ref{K}). Therefore, $M_{\rm new}^{ \lambda \mu \nu}$ 
and $M_{\rm cano}^{ \lambda \mu \nu}$ satisfy the same conservation law and give the same conserved angular 
momentum:
\begin{subequations}
\begin{align}
&\partial_{\lambda} M_{\rm cano}^{\lambda\mu\nu}=\partial_{\lambda} M_{\rm new}^{\lambda\mu\nu}=0,\\
&J^{\mu\nu}=\int{\mathrm{d}}^3 x M_{\rm cano}^{0 \mu \nu }=\int{\mathrm{d}}^3 x\,M_{\rm new}^{0 \mu\nu }.
\end{align}
\end{subequations}

Comparing $M_{\rm new}^{\lambda\mu\nu}$ in Eq. (\ref{Mnew}) with $M_{\rm cano}^{\lambda\mu\nu}$ in Eq. 
(\ref{Mcano}), we see that the last term in Eq. (\ref{Mnew})
can be regarded as an extra spin current. Certainly, the integrated spin ``charge'' is not altered 
by this extra current, which does not contribute to the component $M_{\rm new}^{0ij}$.

\section{Explicit expressions}

For the convenience of future references, we summarize here the explicit expressions. For the scalar field, we have
\begin{subequations}
\begin{align}
{^\phi} T_{\rm new}^{\mu\nu}&=\partial^{\mu}\phi\overleftrightarrow{\partial}^{\nu}\phi=\frac{1}{2}(\partial^\mu\phi
\partial^\nu\phi-\phi\partial^\mu\partial^\nu\phi), \\
{^\phi} M^{\lambda\mu\nu}_{\rm new}&= {^\phi} T^{\lambda\nu} x^{\mu}_{\rm new} -{^\phi} T^{\lambda\mu}_{\rm new} 
x^{\nu} +\frac{1}{2}\phi(g^{\lambda\nu}\partial^{\mu}\phi-g^{\lambda\mu}\partial^{\nu}\phi).
\end{align}
\end{subequations}
Note that ${^\phi} T_{\rm new}^{\mu\nu}$ is still symmetric, yet it is different from the conventional symmetric 
expression
of energy-momentum tensor, which for a scalar field coincides with the canonical expression. More remarkably, the 
scalar field does acquire the extra spin current, though its spin charge is still zero. 

For the Dirac field, the conventional choice of Lagrangian is already zero by the equation of motion, therefore our
``new'' expressions coincide with the conventional canonical expressions: 
\begin{subequations}
\begin{align}
{^\psi} T_{\rm new}^{\mu\nu}&={^\psi} T_{\rm cano}^{\mu\nu}=-i\overline{\psi}\gamma^\mu\overleftrightarrow{\partial}^
\nu\psi 
=\frac{-i}{2}[\overline{\psi}\gamma^\mu\partial^\nu\psi-
(\partial^\nu\overline{\psi})\gamma^\mu\psi],
\\
{^\psi}M^{\lambda\mu\nu}_{\rm new}&={^\psi}M^{\lambda\mu\nu}_{\rm cano}={^\psi}T^{\lambda\nu}_{\rm new}x^{\mu}-
{^\psi}T^{\lambda\mu}_{\rm new}x^{\nu}+\frac{1}{2}\varepsilon^{\lambda\mu\nu\rho}\overline{\psi}\gamma_\rho
\gamma_5\psi .
\end{align}
\end{subequations}
So, for the Dirac field, all we add by our current-correlation analysis is that the symmetric expression of 
energy-momentum tensor is disfavored. 

For the electromagnetic field, we have already put the explicit expression of our new energy-momentum tensor in Eq. 
(\ref{Anew}), and the new angular momentum tensor is
\begin{equation}
{^A}M^{\lambda\mu\nu}_{\rm new}={^A}T^{\lambda\nu}_{\rm new}x^{\mu}-{^A}T^{\lambda\mu}_{\rm new}x^{\nu}+
A^{\mu}F^{\nu\lambda}-A^{\nu}F^{\mu\lambda}
 +\frac{1}{2}(g^{\lambda\nu}A_{\rho} F^{\mu \rho}-g^{\lambda\mu}A_{\rho} F^{\nu \rho}).
\end{equation}
It contains both the ``traditional'' spin current and the extra spin current, but only the former contributes the 
spin charge. 

\section{The quark-gluon system and nucleon structure}\label{nucleon}

For the interacting fields, the Lagrangian necessarily contains higher-than-quadratic terms, therefore, the expression in 
Eq. (\ref{Lquad}) does not hold, and we cannot reach the hyper-canonical form as in Eqs. (\ref{TnewH}) and 
(\ref{MnewH}) for the free fields. 
This is actually reasonable, because generally the individual field of an interacting system can no longer be separately in 
the eigenstate of energy, momentum, or angular momentum, so our current-correlation analysis in quantum measurement
doe not apply. All the guidance we have here is that the N\"other currents should reduce to the hyper-canonical expressions
for the free fields when the coupling constant goes to zero, while in the presence of interaction,  they must satisfy the same 
conservation laws and give the same conserved charges as the conventional expressions do. 
Let us tentatively modify an interacting Lagrangian with the same surface term as in Eq. (\ref{surface}). [Certainly, in the 
presence of interaction we can no longer reach Eq. (\ref{Lnew}).]
For example, take the standard expression of QCD Lagrangian, 
\begin{eqnarray}
{\mathscr L}_{\rm QCD}&=&\overline{\psi}({\rm i}\gamma^{\mu}\overleftrightarrow{\partial}_{\mu}-m)\psi-\dfrac{1}{4}F_{\mu\nu}^{a} F^{\mu\nu a}+g\overline{\psi}\gamma^{\mu}t^{a}\psi A_{\mu}^{a} \nonumber \\
 &\equiv&{\mathscr L}_{\rm q}+{\mathscr L}_{\rm g}+{\mathscr L}_{\rm qg},
\end{eqnarray}
we get
\begin{eqnarray} 
\widetilde{{\mathscr L}}_{\rm QCD}\equiv {\mathscr L}_{\rm QCD}-
\dfrac{1}{2}\partial_{\mu}(\dfrac{\partial {\mathscr L}_{\rm QCD}}{\partial (\partial_{\mu}\phi_a)}\phi_a)
={\mathscr  L}_{\rm q}+\widetilde{{\mathscr L}}_{\rm g}+{\mathscr L}_{\rm qg},
\end{eqnarray}
where the modified gluon part is
\begin{equation}
\widetilde{{\mathscr L}}_{\rm g}=\dfrac{1}{2}A_{\nu}^{a} \partial_{\mu}F^{\mu\nu a}-
\dfrac{1}{4}g f^{abc}F^{\mu\nu a}A_{\mu}^{b} A_{\nu}^{c}. 
\end{equation}

With this Lagrangian, the new energy-momentum and angular momentum tensors are derived to be
\begin{subequations}\label{QCD}
\begin{align}
{^{\rm QCD}}{\mathcal T}_{\rm new}^{\mu\nu}=&-i\overline{\psi}\gamma^{\mu}\overleftrightarrow{\partial}^{\nu}\psi+F^{\mu\rho a}\overleftrightarrow{\partial}^{\nu}A_{\rho}^{a}+g^{\mu\nu}\widetilde{\mathscr L}_{\rm QCD}, 
\label{TQCD}
\\
{^{\rm QCD}}{\mathcal M}_{\rm new}^{\lambda \mu \nu}=&- i \overline{\psi}\gamma^{\lambda}x^{[\mu} 
\overleftrightarrow{\partial}^{\nu]} \psi
\longrightarrow {\mathcal M}_{\rm q,orbital}^{\lambda \mu \nu} 
\nonumber \\
&+F^{\lambda\rho a}x^{[\mu} \overleftrightarrow{\partial}^{\nu]} A_{\rho}^a
\longrightarrow {\mathcal M}_{\rm g,orbital}^{\lambda \mu \nu}
\nonumber \\
&+\frac{1}{2}\varepsilon^{\mu \nu\lambda\sigma}\overline{\psi}\gamma_{\sigma}\gamma^{5}\psi\longrightarrow {\mathcal M}_{\rm q,spin}^{\lambda \mu \nu}
\nonumber \\
&+F^{\lambda[\mu}_{a}A^{\nu]}_{a}+\frac{1}{2}g^{\lambda[\nu}F^{\mu]\rho a}A_{\rho}^{a}\longrightarrow {\mathcal M}_{\rm g,spin}^{\lambda \mu \nu}
\nonumber \\
&+x^{[\mu} g^{\nu]\lambda}\widetilde{\mathscr L}_{\rm QCD}\longrightarrow{\mathcal M}_{\rm boost}^{\lambda \mu \nu},
\label{MQCD}
\end{align}
\end{subequations}
where the indices in square brackets are to be anti-symmetrized. They can indeed reduce to the free-field
expressions in the absence of interaction. Moreover, for the momentum density, the Lagrangian term in 
${^{\rm QCD}}{\mathcal T}_{\rm new}^{0i}$ drops out, and for the angular momentum density, the boost term in ${^{\rm QCD}}
{\mathcal M}_{\rm new}^{0ij}$ drops out, so we get exactly the same expression as if quark and gluon exist separately. 

Like the canonical expressions in gauge theory, our new expressions of energy-momentum tensor and angular momentum tensor
are naively gauge-dependent. 
To achieve gauge-invariance, one can employ the method as discussed in Refs. \cite{Lead14, Chen08, Chen09}, by separating 
the gauge field into a physical part and a pure-gauge part:  
\begin{equation}
A^{\mu}=A^{\mu}_{pure}+A^{\mu}_{phys}.
\end{equation}
With this method, Eqs. (\ref{QCD}) can be upgraded to be gauge-invariant: 
\begin{subequations}\label{QCD'}
\begin{align}
{^{\rm QCD}}{\mathrm T}_{\rm new}^{\mu\nu}=&-i\overline{\psi}\gamma^{\mu}\overleftrightarrow{D}^{\nu}_{pure}\psi
+2{\rm Tr}\big[F^{\mu\rho}\overleftrightarrow{\mathscr D}^{\nu}_{pure}A_{\rho}^{phys}\big]
+g^{\mu\nu}\widetilde{\mathscr L}_{\rm QCD}^{phys}, 
\\
{^{\rm QCD}}{\mathrm M}_{\rm new}^{\mu \nu\lambda}=&- i \overline{\psi}\gamma^{\lambda}x^{[\mu} 
\overleftrightarrow{D}^{\nu]}_{pure} \psi\longrightarrow {\mathrm M}_{\rm q,orbital}^{\lambda \mu \nu}
\nonumber\\
&+2{\rm Tr}\big[F^{\lambda\rho}x^{[\mu} \overleftrightarrow{\mathscr D}^{\nu]}_{pure}A_{\rho}^{phys}\big]
\longrightarrow {\mathrm M}_{\rm g,orbital}^{\lambda \mu \nu}
\nonumber\\
&+\frac{1}{2}\varepsilon^{\mu \nu\lambda\sigma}\overline{\psi}\gamma_{\sigma}\gamma^{5}\psi\longrightarrow {\mathrm M}_{\rm q,spin}^{\lambda \mu \nu}
\nonumber\\
&+2{\rm Tr}\big[F^{\lambda[\mu}A^{\nu]}_{pyhs}+\frac{1}{2}g^{\lambda[\nu}F^{\mu]\rho}A_{\rho}^{phys}\big]\longrightarrow {\mathrm M}_{\rm g,spin}^{\lambda \mu \nu}
\nonumber\\
&+x^{[\mu} g^{\nu]\lambda}\widetilde{\mathscr L}_{\rm QCD}^{phys}\longrightarrow {\mathrm M}_{\rm boost}^{\lambda \mu \nu},
\end{align}
\end{subequations}
where the gauge-covariant derivative is constructed with the pure-gauge field $A^{\mu}_{pure}$, and the gauge-invariant 
modified Lagrangian is 
\begin{equation}
\widetilde{\mathscr L}_{\rm QCD}^{phys}={\mathscr L}_{\rm QCD}-\dfrac{1}{2}\partial_{\mu}(\dfrac{\partial {\mathscr L}_{\rm QCD}}{\partial(\partial_{\mu}A_{\rho})}A_{\rho}^{phys})={\mathscr L}_{\rm QCD}+\partial_{\mu}{\rm Tr}\big[F^{\mu\rho}A_{\rho}^{phys}\big] .  
\end{equation}

If we consider the integrated spin, momentum, or orbital angular momentum of quarks and gluons, 
each term in Eqs. (\ref{QCD'}) coincide with the corresponding 
term in Ref. \cite{Chen09}, this justified the gauge-invariant canonical decomposition of nucleon momentum and spin in 
Ref. \cite{Chen09}. Note, however, that if the flux densities of momentum and angular momentum are considered, 
Eqs. (\ref{QCD}) and (\ref{QCD'}) differ from the canonical expressions. 

\section{Discussion}
In this paper, we employed the particle-wave duality in quantum mechanics to set a first-principle constraint on the 
expressions of energy-momentum tensor and angular momentum tensor. 
It should be reminded, nevertheless, that although the foundation of our whole discussion --- 
the mutual conservation of multiple physical quantities during a quantum measurement, or the simultaneous 
localization of multiple quantities to the same spot during the collapse of a quantum wave --- sounds very 
natural and reasonable, it yet has never been actually verified by experiment, as far as we know. 
We encourage that such a fundamental, important, and taken-for-granted quantum property be tested.  

It is interesting to note that in Ref. \cite{Came12,Blio13}, the dual symmetry between electric field $\vec E$ and 
magnetic field $\vec B$ is employed to derive alternative expressions for the energy-momentum and 
angular momentum tensors of the electromagnetic field. We checked that their expressions also satisfy the 
current correlation as we discussed here. Certainly, the method based on $\vec E$-$\vec B$ dual-symmetry cannot be 
applied to a general field.  

The most important use of energy-momentum tensor is to generate gravity. In another paper \cite{Lei17},
we will construct a theory with our new energy-momentum tensor and spin tensor acting as the source of 
gravity and torsion.  

This work is supported by the China NSF via Grants No. 11535005 and No. 11275077.

\end{document}